# An Adaptive Quasi Harmonic Broadcasting Scheme with Optimal Bandwidth Requirement

Farzana Afrin
Department of Computer Science
American International Univrsity-Bangladesh
Dhaka, Bangladesh
farzana@aiub.edu

Mohammad Saiedur Rahaman
Department of Computer Science
American International Univrsity-Bangladesh
Dhaka, Bangladesh
saied@aiub.edu

*Abstract*- The aim of Harmonic Broadcasting protocol is to reduce the bandwidth usage in video-on-demand service where a video is divided into some equal sized segments and every segment is repeatedly transmitted over a number of channels that follows harmonic series for channel bandwidth assignment. As the bandwidth of channels differs from each other and users can join at any time to these multicast channels, they may experience a synchronization problem between download and playback. To deal with this issue, some schemes have been proposed, however, at the cost of additional or wastage of bandwidth or sudden extreme bandwidth requirement. In this paper we present an adaptive quasi harmonic broadcasting scheme (AQHB) which delivers all data segment on time that is the download and playback synchronization problem is eliminated while keeping the bandwidth consumption as same as traditional harmonic broadcasting scheme without cost of any additional or wastage of bandwidth. It also ensures the video server not to increase the channel bandwidth suddenly that is, also eliminates the sudden buffer requirement at the client side. We present several analytical results to exhibit the efficiency of our proposed broadcasting scheme over the existing ones.

## I. INTRODUCTION

Traditional VOD system [1]-[6] [9] [11-12] works like that after receiving client requests for videos, the VOD system provides the clients with a dedicated server channel and the clients are served instantly. In this system, server channels are specially controlled by the clients because the server is bound to serve the video immediately after getting a request from client end and thus this type of service is widely known as true video on demand (TVOD) service. Here the server load depends on the rate of viewer's arrival. The higher rate of viewer's arrival linearly increases the load on the server side which is a serious bottleneck at the server-side of TVOD system. Additionally, when a VOD server is running at the remote place, the service provider must think about the high cost of using dedicated links to serve the client requests. Again, these clients may experience from congestion, delay and jitter during transmission as they are getting the service from a remote place. So, this system cannot be best suited with large number of client requests and thus it is said that TVOD systems are not scalable. So, researchers are continuously trying [10] to reduce the load on VOD server eliminating the effects of congestion, transmission delay and jitter on the client side at the same time reducing the broadcasting cost.

Near video on demand (NVOD) is another VOD service where the main concern is to reduce the bandwidth requirement at the server side. This is achieved by batching the client requests over a specified interval and forming a viewers group requesting the same video to share a single video stream. The consequence of these techniques is a short waiting time on the client side. Hence here the viewers cannot watch the video immediately. This is the grounds of calling these schemes near video on demand.

Harmonic Broadcasting is one of the pioneer NVOD schemes. Here a video is divided in to some equal sized segments and then broadcasted in to some channels. These channels follow harmonic series for channel bandwidth allocation. But the major pitfall of traditional harmonic scheme is that it cannot deliver some segments when necessary and thus leads to a problem called download and playback synchronization problem. To overcome this problem three broadcasting schemes have been proposed in [4] and [12] known as *Cautious harmonic broadcasting protocol (CHB)*, *Quasi-harmonic broadcasting protocol (QHB)* and *Adaptive Harmonic broadcasting (AHB)*. First two schemes remove the synchronization problem but introduce the cost of additional bandwidth requirement by assigning additional channel bandwidth or by transmitting some extra segments respectively and the third scheme introduce sudden bandwidth increment in transmission channels compared to traditional harmonic broadcasting. In this research we suggest an adaptive quasi harmonic broadcasting scheme that eliminates the download and playback synchronization problem of traditional harmonic broadcasting protocol while keeping the bandwidth consumption exactly same as the traditional harmonic broadcasting protocol. It also ensures no sudden bandwidth requirement in transmission channels. The proposed protocol adapts itself with the problems of download and playback synchronization, additional channel bandwidth assignment, redundant segment transmission and sudden bandwidth increment in transmission channels and that's the reason why we call it the Adaptive quasi harmonic broadcasting (AQHB) scheme.

The organization of the paper is as follows. Section 2 highlights some related works. Section 3 presents the proposed adaptive quasi harmonic broadcasting scheme (AQHB) and its features. We have shown some analytical results and comparison of our scheme with traditional *Harmonic Broadcasting Protocol* (HB), *Cautious harmonic broadcasting protocol (CHB)*, *Quasi-harmonic broadcasting protocol (QHB)* and *Adaptive Harmonic broadcasting (AHB)* in Section 4. Finally Section 5 concludes the paper.

## II. RELATED WORKS

### 2.1 Harmonic Broadcasting Scheme

*Harmonic broadcasting* (HB) [1]-[3],[5]-[6], [9], [12] protocol is a pioneer NVOD scheme intended for efficient delivery of on demand video data that are concurrently watched by a number of clients those who requested for the identical video. Here, a video of size $S$ and length $T$ is divided into $N$ equal segments and each segment is then periodically broadcasted on a dedicated channel. However, successive transmission channels follow harmonic series in decreasing order for bandwidth assignment. Assuming a video playback rate $b$, the first video segment $S_1$ is broadcasted through first channel at a bandwidth $b$, the second channel broadcasts second segment $S_2$ at bandwidth $b/2$, the third channel broadcasts third segment $S_3$ at bandwidth $b/3$ and so on (Figure 1). The clients are able to download the segments from all the channels simultaneously and the segments or sub segments that are downloaded in advance are stored at a local storage at client side. The maximum client waiting time in HB scheme is

$$w = \Delta t = \frac{T}{N} \tag{1}$$

and the server bandwidth required in HB scheme in a particular time slot of length $\Delta t$ is given by:

$$B_{HB} = b \times \sum_{i=1}^{N} 1/i = b \times H(N) \tag{2}$$

Here, $N$ is the total number of video data segments and $H(N)$ is the sum of harmonic series over integer $N$.

One major problem that HB scheme sometimes experiences is the *download and playback synchronization problem* [5]-[6]. For example, in Figure 1, a client who starts at $t_2$ will finish the download and playback of the first segment $S_1$ at $t_3$ at bandwidth $b$. At that time the client has only the second half of the second segment $S_2$ in its local buffer. Now, if the client starts downloading the first half of the second segment at $t_3$ at bandwidth $b/2$, it will not be possible to play the video at bandwidth $b$ and thus the download and playback synchronization problem occurs. To resolve this problem, a simple display delay equal to the size of the first segment can be imposed however, at the cost of making the client waiting in the middle of watching a video may not be an ideal and acceptable situation.

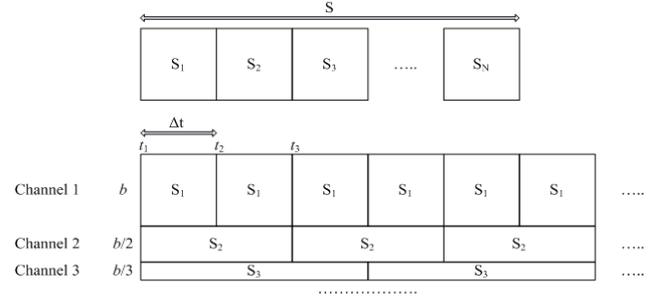

Figure 1: The HB protocol with video $V$ divided into segments $S_1$, $S_2$,…,$S_N$ and the segments are transmitted on different channels at decreasing bandwidth following the harmonic series. In a time slot $\Delta t$ the total server bandwidth requirement equals $b \times H(N)$.

### 2.2 Cautious Harmonic Broadcasting Scheme

To resolve the abovementioned synchronization problem, Paris, Carter and Long proposed another harmonic broadcasting based protocol called *Cautious harmonic broadcasting* (CHB) [4], but it costs some additional bandwidth. Assuming a video of size S divided into $N$ segments, the CHB scheme (Figure 2) assigns full bandwidth $b$ to its first two transmission channels. First channel repeatedly broadcasts first segment $S_1$ while second channel broadcasts second and third segments $S_2$ and $S_3$ by rotation. Other successive segments are broadcasted on separate broadcasting channels at decreasing bandwidth partly following the harmonic series starting from $b/3$. In particular, segment $S_i$ (where, $4 \leq i \leq N$) is broadcasted at channel $i-1$ at bandwidth $b/(i-1)$. This new arrangement of video segments ensures that the customer will either receive a video segment at full bandwidth when it is needed, or have the entire segment already in its local buffer before it is needed. So, the bandwidth requirement of the CHB protocol is given by:

$$\begin{aligned} B_{CHB} &= (b+b) + \sum_{i=3}^{N-1} b/i \\ &= 2b + \sum_{i=1}^{N-1} b/i - b - b/2 \\ &= b/2 + b \times H(N-1) \end{aligned} \tag{3}$$

Here, $b$ is the rate of video playback; $N$ is the total number of video data segments and $H(N-1)$ is sum of the harmonic series over $N-1$.

Although CHB solves the synchronization problem, a major downside of this scheme is that first two channels serves the video segments at full bandwidth and thus this technique leads to a higher bandwidth consumption than traditional HB as

$$\begin{aligned} B_{CHB} &= b/2 + b \times H(N-1) \\ &= b/2 + b \times (H(N) - 1/N) \\ &= b \times H(N) + (b/2 - b/N) \\ &= B_{HB} + (b/2 - b/N) \end{aligned} \tag{4}$$

and $(b/2 - b/N) > 0$ for $N > 2$.

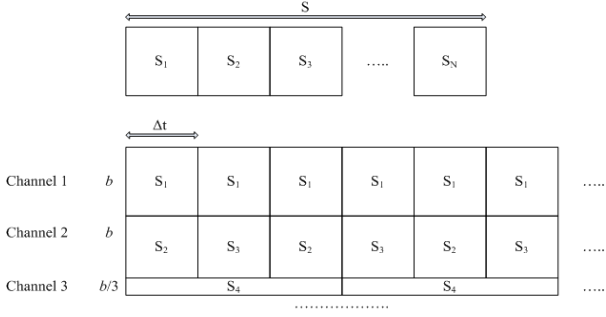

Figure 2: Delivery of video segments in CHB. Assuming *N* segments, the first channel broadcasts segment $S_1$ at playback rate *b*, the second channel repeatedly broadcasts video segments $S_2$ and $S_3$ at bandwidth *b* whereas the third channel broadcasts segment $S_4$ at bandwidth $b/3$ and in general segment *i* ($4 \leq i \leq N$) is served with bandwidth $b/(i–1)$.

## 2.3 Quasi-harmonic broadcasting

Like other harmonic VOD broadcasting protocols, *Quasi-harmonic broadcasting* (QHB) also breaks up each video into *N* equal segments and broadcasts the first segment repeatedly through the first channel [4]. But unlike others, the QHB protocol (Figure 3) divides each segment *i* (where, $2 \leq i \leq N$) into $im - 1$ fragments (Here, *m* is any fragmentation parameter), and the client receives *m* fragments from *i*-th channel per time slot (a time slot equals the duration of first segment, $S_1$). If each time slot is divided into *m* equally sized sub slots, then a client receives a single fragment during each sub slot from a specific channel. The most vital issue regarding QHB is the placement of fragments in each channel. For any channel *i* in QHB scheme, the last sub slot of each time slot is used to transmit the first $i - 1$ fragments of $S_i$ in order. The rest of the sub slots transmit the other $i \times (m - 1)$ fragments in such a manner so that the *k-th* sub slot of slot *j* is used to transmit fragment $(ik + j - 1) mod(i(m - 1)) + i$.

This ordering scheme of fragment transmission introduces a problem of repetitive transmission of the first fragment of any segments twice and thus for timely delivery of fragments to avoid download and playback synchronization problem, this scheme needs to allocate a channel bandwidth of $bm/im - 1$ for $2 \leq i \leq N$ which is greater compared to traditional HB.

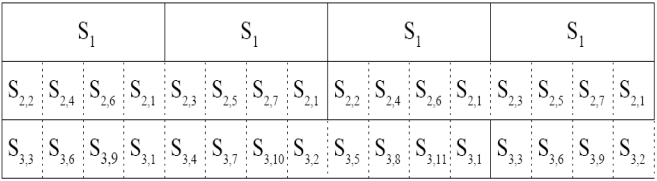

Figure 3: An illustration of delivery of segments/fragments in the first three channels for a *video* in QHB scheme with *m*=4.

## 2.4 Adaptive Harmonic Broadcasting

Assuming a video of length *T*, and the playback rate of the video *b* (i.e. the video size is $S = T \times b$, the AHB scheme (Figure 4) involves the following steps [12]:

i. For any positive integer *N*, the video is equally divided in to *N* segments. Suppose, $S_k$ is the *k*-th segment of the video (where $1 \leq k \leq N$). So, the concatenation of all segments, in increasing order of segment numbers, constitutes the video as a whole i.e. $S = S_1 \bullet S_1 \bullet \cdots \bullet S_N$.

ii. The *k*-th segment of the video, $S_k$ is divided equally into *k* fragment(s) and hence $S_k = S_k^1 \bullet S_k^2 \bullet \cdots \bullet S_k^k$ where, $S_k^i$ indicates the *i*-th fragment of the *k*-th segment.

iii. The *k*-fragments of segment $S_k$ are arranged on channel $C_k$ which is a *variable bit rate (VBR)* channel. For normal fragments, the bandwidth assigned to $C_k$ is $b/k$, whereas on the time slots that serves problematic fragments (i.e. fragments with download and playback synchronization problem), bandwidth allocation for $C_k$ is *b* (Figure 4) for a duration of $\Delta t/k$ where $\Delta t$ denotes the duration of segment $S_1$. In AHB scheme, a problematic fragment *i* for segment *k*, $S_k^i$ starts at $kt + (i - 1)$ where, $k > 1$, $i \in \{1,2, \ldots, k - 1\}$ and $t \in \{1,2, \ldots, \infty\}$.

iv. Within channel $C_k$ the *k* fragment(s) of Segment $S_k$ will be broadcasted periodically (Figure 4).

The AHB scheme eliminates the download and playback synchronization problem present in HB. As shown in Figure 4, a client who starts at $t_2$, will finish the playback of the first segment at the start of $t_3$. At that time the client has only the second fragment of the second segment in its local buffer. Using HB scheme (Figure 1), the client starts downloading the first fragment of the second segment at time slot two at bandwidth $b/2$ and will not be able to playback it at rate *b*. But using AHB protocol, the client can start downloading the first fragment of the second segment at time slot two as it is served with bandwidth *b* and can playback at the same rate and hence the client will not experience any lack of synchronization between download and playback and will not experience any waiting in the middle of watching a video.

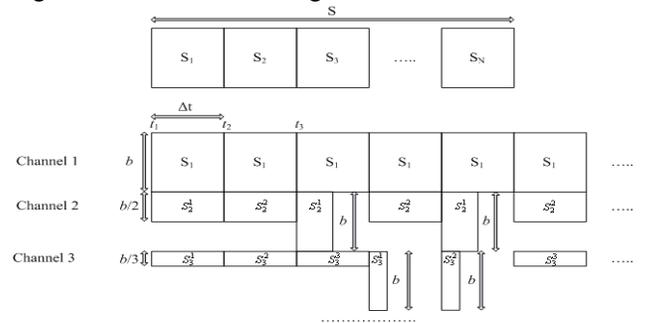

Figure 4: Fragments delivery in the AHB scheme with *N*=3 segments. The problematic fragments on channel 2, and 3 are served at bandwidth *b* to ensure timely delivery of video fragments.

## III. PROPOSED SCHEME AND IT'S FEATURES

Our proposed scheme aims to

i. Deal with repeated transmission problem of fragments and additional channel bandwidth allocation problem exists in Quasi Harmonic Broadcasting scheme.

ii. Keeps the property of QHB that ensures timely delivery of segments leading to synchronization between download and playback so that a client needs not to wait in the middle of watching a video.
iii. Consume the same amount of bandwidth as the HB scheme.

Our scheme also proposes a modification to the existing Adaptive Harmonic Broadcasting (AHB) to guarantee:

i. There will be no sudden increment of bandwidth in transmission channels.
ii. There will be no download and playback synchronization problem.
iii. Bandwidth consumption will be as same as the traditional HB scheme even in worst cases.

As the proposed scheme adapts itself to the synchronization problem without increasing or wasting the bandwidth requirement we prefer to name it Adaptive Quasi Harmonic Broadcasting (AQHB) scheme. In the remainder of this section, we first present proposed AQHB scheme and then some of its features.

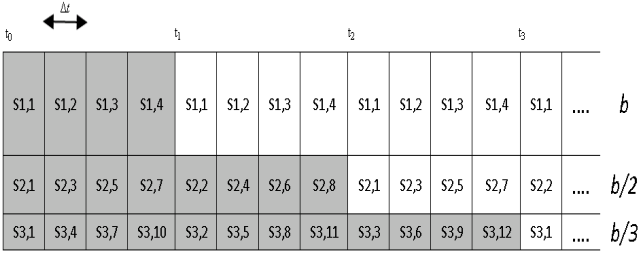

Figure 5: Delivery of fragments in the proposed AQHB scheme with N=3 segments.

### 3.1 Adaptive Quasi Harmonic broadcasting

The operation of the AQHB starts with assuming that the size of the video is $S$ with length $T$ and the video playback rate is $b$. The proposed AQHB scheme (Figure 4) involves the following steps:

i. The video $S$ is equally partitioned in to $N$ segments, where $N$ is a positive integer. Suppose $S_i$ is the $i$-th segment of the video where $1 \leq i \leq N$. So, the concatenation of all the segments constitutes the whole video i.e. $S = S_1 \bullet S_2 \bullet \cdots \bullet S_N$.

ii. Assuming each time slot is equal to the playback time of a segment, each time slot is then divided into $m$ equal sub-slots and client will receive fragments from different channels during each sub-slot simultaneously. So, a client will only wait to start to play the video till first fragment of the first segment being downloaded.

iii. The $i$-th segment of the video, $S_i$ is equally divided into $im$ fragment(s) to be transmitted and thus $S_i = Si,1 \bullet Si,2 \bullet Si,3 \bullet ... \bullet Si,im$ where, $Si,im$ denotes the $im$-th fragment of the $i$-th segment.

iv. Before start transmission of fragments of any segment $i$ the AQHB scheme creates a matrix $M_{r \times c}$ with dimensions $r \times c$. Here $r = i$ $and$ $c = m$. The matrix is filled up using the following algorithm (Figure 5(a)):

*for (row = 1 to i)*

*for(col = 1 to m)*

*{*

*Find the Row index, RI*

*Calculate $j = i \times (col - 1) + RI$*

*Fill $M_{row,col}$ with fragment $Si,j$*

*}*

Figure 5(a): Matrix Generation Algorithm

v. Within channel $i$ the fragment(s) of Segment $S_i$ from matrix $M_{r \times c}$ will be broadcasted row wise from left to right periodically (Figure 5(b)) i.e. for the initial value of $RI = 1$, the fragments of the first row of the matrix will be transmitted from left to right, for the value of $RI = 2$, the fragments of the second row of the matrix will be transmitted from left to right and so on. The value of $RI$ comes down to 1 immediately when it finishes the transmission of the row with maximum $RI$ value which is equal to $i$ and the whole process continues till the server is alive.

| | | | | |
|---|---|---|---|---|
| $t_0$ | S3,1 | S3,4 | S3,7 | S3,10 |
| $t_1$ | S3,2 | S3,5 | S3,8 | S3,11 |
| $t_2$ | S3,3 | S3,6 | S3,9 | S3,12 |

Figure 5(b): Fragment distribution for fourth segment ($S_3$) in the matrix.

Note: $S_3$ will be broadcasted at third channel ($C_3$) through three consecutive time slot starts at $t_0, t_1,$ and $t_2$ at bandwidth $b/3$ and then repeats again.

The proposed AQHB scheme eliminates the problem regarding download and playback synchronization in HB and sudden channel increment in AHB. As shown in Figure 4, a client who starts at $t_1$, will start playback the first fragment $S1,1$ at $t_1 + \Delta t$ and finish the playback of the first segment at $t_2 + \Delta t$. In the next $\Delta t$ duration of time the client need to play $S2,1$ and $S2,2$ which is already in its local buffer. Using HB scheme (Figure 1), the client starts downloading the first half of the second segment at time slot two at bandwidth $b/2$ and will not be able to playback at rate $b$. Using AHB scheme (Figure 4) client need to serve the missing fragment through a patch channel with additional bandwidth. But using our proposed AQHB scheme (Figure 5) the client already have $S2,1$ and

$S_{2,2}$ to be played at next Δt after $t_2$ + Δt and can playback at the same rate and the client will not face any lack of synchronization between download and playback. So, we can claim that our proposed AQHB scheme delivers all data segment on time.

### 3.2 Features of AQHB

Alongside solving the synchronization problem AQHB has some salient features as presented bellow is this section.

#### 1) Initial Client Waiting Time
The initial waiting time in the worst case is proportional to the duration of segment $S_1$ i.e. $T/N$ where a video of length $T$ is partitioned into $N$ segments.

#### 2) Bandwidth Consumption in a Particular Time Slot
The bandwidth consumption $B_{AQHB}(k)$ in AQHB scheme (Figure 5) for a single channel $k$ in a time slot is

$$B_{AQHB}(k) = b/k \quad (5)$$

So after combining all the channels in general, it becomes

$$B_{AQHB} = \sum_{k=1}^{N} B_{AQHB}(k) = \sum_{k=1}^{N} b/k \quad (6)$$
$$= b \times \sum_{k=1}^{N} 1/k = b \times H(N) = B_{HB}$$

The last equality in the previous equation follows from (2). It can thus be observed that AQHB schemes consumes the same bandwidth as HB scheme while solving the synchronization problem. Note that CHB consumes more bandwidth than HB whereas QHB wastes some bandwidth and AHB introduces sudden bandwidth in channels thus making AQHB superior to them.

#### 3) Storage Requirement
In Figure 4, we observe that a client joining at any time slot needs to download from all the channels. But the storage requirement is same as HB scheme and thus again makes AQHB superior to all other variants of HB scheme.

#### 4) Waiting Time for Discontinued Fragment
As there remains no discontinued frame, a client does not need to wait in the middle of watching a video to eliminate the problem as in HB scheme. Hence AQHB provides zero waiting time for discontinued fragments.

## IV. ANALYTICAL RESULTS AND DISCUSSION

### 4.1 Bandwidth Consumption in a Particular Time Slot

From Figure 6, we see that in a specific time slot, the bandwidth consumption of proposed AQHB scheme is equal to the HB scheme whereas less than QHB and CHB scheme. Again, in some cases, AHB scheme needs to serve all the channels at full bandwidth whereas it consumes no sever bandwidth in some cases. But our proposed scheme needs not to serve all the channels at full bandwidth in any case or never keeps the channel idle. That is AQHB requires a static bandwidth equal to the traditional HB scheme and it is the acceptable solution to bandwidth consumption compare to any variants of HB scheme.

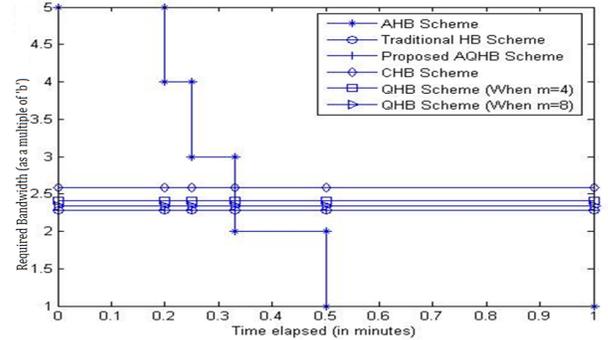

Figure 6: Comparison of bandwidth consumption between HB, QHB, CHB, AHB and AQHB scheme.

### 4.2 Waiting Time for Discontinued Fragments

Figure 7 shows the waiting time for lack of download and playback synchronization in traditional HB scheme, QHB scheme, CHB scheme, AHB scheme and proposed AQHB scheme. It is zero for CHB, QHB, AHB and AQHB scheme because there is no lack of synchronization in these schemes. But in traditional HB scheme the waiting time is inversely proportional to the number of segments. As the number of video segments increases in HB, the waiting time for lack of synchronization decreases. The simulation shows for a 120 minutes video that the video with 1 segment experiences a waiting time of 120 minutes at the client side, whereas for the video with 2,3,4, and 5 segments the waiting time at the client end is reduced to 60 minutes, 40 minutes, 30 minutes and 20 minutes respectively.

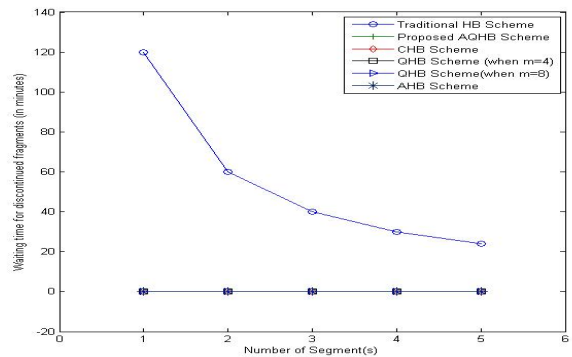

Figure 7: Comparison of client waiting time between HB, QHB, CHB, AHB and AQHB scheme for discontinued segments in a 120 minutes video with number of segments = 5.

### 4.3 Comparison with Existing HB Scheme and its Variants

In this section, we provide a comparison among the proposed AQHB scheme AHB scheme, the traditional HB scheme, Quasi Harmonic Broadcasting (QHB) scheme and Cautious

Harmonic Broadcasting (CHB) scheme considering initial client waiting time [7][8], buffer requirement etc. From previous section, we can see that the proposed scheme solve the download and playback synchronization problem without additional average bandwidth as done in CHB and QHB. The following table (Table I) provides a comparative picture between HB, CHB, QHB, AHB and AQHBschemes.

## V. CONCLUSION

In this paper we have presented an adaptive QHB scheme (AQHB) that resolves the problem of intermediate client waiting time by synchronizing between video download and playback rates. Compared to traditional HB scheme it consumes the same amount of bandwidth whereas other modifications consume more bandwidth (e.g. CHB and QHB). The worst case bandwidth requirement of AHB, however, is higher than others. Another common type of synchronization problem arises when number of channels change as it creates a correspondence problem between segment sizes [9]. While this problem is dealt with other VOD broadcasting schemes, to our knowledge it is still not explored for HB scheme. In future we aim to extend the AQHB scheme to deal with the channel transition problem.

TABLE I: COMPARISON OF AQHB SCHEME WITH TRADITIONAL HB, CHB, QHB AND AHB SCHEME

| Criteria | AQHB vs HB scheme | AQHB vs CHB scheme | AQHB vs QHB scheme | AQHB vs AHB scheme |
| --- | --- | --- | --- | --- |
| Bandwidth consumption in a particular time slot | Equal | Less in AQHB | Less in AQHB | Equal with AHB( in average case) but less than AHB(in worst case) |
| Initial client waiting time | Equal | Equal | Equal | Equal |
| Storage requirement | Equal | Less in AQHB | Less in AQHB | Equal with AHB( in average case) but less than AHB(in worst case) |
| Waiting time for discontinued fragments | No waiting time in AQHB | No waiting time in AQHB and CHB | No waiting time in AQHB and QHB | No waiting time in AQHB and AHB |
| Download and playback synchronization | AQHB provides but HB does not | AQHB and CHB provides | AQHB and QHB provides | AQHB and AHB provides |